# Demonstration of an optical polarization magnifier with low birefringence

M. Lintz[a], J. Guéna, M.-A. Bouchiat and D. Chauvat[b]

Laboratoire Kastler Brossel[c] and Fédération de Recherche[d]

Ecole Normale Supérieure, Département de Physique

24 rue Lhomond

75231 Paris cedex 05, France

**Abstract**

**In any polarimetric measurement technique, enhancing the laser polarization change of a laser beam before it reaches the analyzer can help in improving the sensitivity. Enhancement of a small polarization rotation can be performed using an optical component having a large linear dichroism, the enhancement factor being equal to the square root of the ratio of the two transmission factors. A pile of parallel plates at Brewster incidence looks appropriate for realizing such a polarization magnifier. In this paper, we address the problem raised by the interference in the plates and between the plates, which affects the measurement by giving rise to birefringence. We demonstrate that wedged plates provide a convenient and efficient way to avoid this interference. We have implemented and characterized devices with 4 and 6 wedged plates at Brewster incidence which have led to a decisive improvement of the signal-to-noise ratio in our ongoing Parity Violation measurement.**

## I. INTRODUCTION

Measuring the state of polarization of a light beam lies at the heart of many fundamental physics experiments, such as for instance parity violation measurements in atoms[1-4] or the detection of vacuum magnetic birefringence[5], as well as applied physics measurement techniques like magnetometry[6-8] and ellipsometry[9]. In these experiments, the quantum (shot-noise) ultimate limit to the measurement of a change in the linear polarization of a laser beam is often desired. However, extra sources of optical noise are usually present and can severely limit the sensitivity of the measurement. In this case, enhancing the polarization change before it reaches the polarimetric detection system helps in improving the sensitivity. Enhancement of a polarization rotation can be achieved using an additional optical component with a large linear dichroism exhibiting different intensity transmission factors $T_x$ and $T_y$ along its two optical axes. A small rotation $\psi$ of the incident beam polarization is then transformed into a much larger one $k\psi$ of the output beam polarization, where $k=\sqrt{T_y/T_x}$ can be several times unity.

So far, two realizations of such a polarization magnifier have been proposed[10,11]. In Ref.[10], the dichroic component is a plate (or a beam splitter cube) with a multilayer coating, which is convenient to use, being a single optical component. But it also has some disadvantages: it operates in a limited wavelength range, its magnification factor $k$ is a rapidly varying function of the incidence angle, and, most importantly, it exhibits birefringence, which is also incidence-dependent. This birefringence, or the phase difference $\Delta\phi$ between the two components of the electric field along the optical axes x and y of the magnifier, is harmful since *i*) it couples to a possible ellipticity $\xi$ in the polarization of the analyzed beam giving rise to an unwanted contribution to the measured linear polarization signal, and *ii*) it results in a loss in the effective magnification factor. To be explicit, the tilt $k\psi$ of the output beam polarization becomes $k(\psi\cos(\Delta\phi)+(\xi/2)\sin(\Delta\phi))$ in the presence of birefringence. In Ref.[10], birefringence was shown to be incidence-dependent, with values up to several tens of degrees.

To overcome the birefringence problem, one can use a pile of $N$ uncoated glass plates at Brewster incidence[11]. This device, which attenuates the polarization perpendicular (⊥) to the plane of incidence and completely transmits the parallel (//) one, is essentially wavelength-independent, and allows easy change of the magnification factor $k$ by changing $N$. At first sight, one might expect that $\Delta\phi=0$ since for a single pass propagation through a parallel plate, the phase shift is the same for // and ⊥ polarizations. However, as one can verify experimentally, a laser beam with linear polarization out of the // or ⊥ directions acquires an ellipticity after crossing such a parallel plate at oblique incidence, showing that $\Delta\phi\neq0$. This effect arises from *interferences* inside the plate, as mentioned in Ref.[12]. This is obvious in the case of exact Brewster incidence where only one beam is transmitted for the // component, while *multiple reflections* occur for the ⊥ polarization component, leading to multiple transmitted beams with different phase shifts due to their propagation inside the plate. Those beams interfere with each other and the phase of the overall ⊥ amplitude then differs from that of the // amplitude. For clarity, we shall call this pseudo-birefringence an "interferential birefringence" in contrast to the "natural birefringence" due to stress. Note that this interference effect is also at the origin of the phase $\Delta\phi$ mentioned in Ref.[10] for the multilayer coated device. A derivation of $\Delta\phi$ due to those interferences is given in Appendix A where it is shown that a phase difference $\Delta\phi$ of up to 7.2° is expected for a single plate at Brewster incidence. As stated above, such a large birefringence is detrimental for precise polarimetric measurements.

Let us also note that an interferential birefringence exists as long as the intensity reflection coefficients of a single side of the plate $R_{//}$ and $R_\perp$ are different, which is required for polarization rotation magnification.

As usual for an interference effect, $\Delta\phi$ varies with temperature, wavelength and incident angle, a serious drawback. However, no spurious interferential birefringence will appear if one is able to separate spatially and eliminate the multiply reflected beams. This could be done by using thick plates but the system of $N$ plates would then be cumbersome. Alternatively, multiple beams can easily be separated by using wedged plates. This paper demonstrates a practical polarization magnifier with low overall birefringence using several wedged silica plates near Brewster incidence.

## II. IMPLEMENTATION OF THE POLARIZATION MAGNIFIER

For our application[13], an enhancement factor $k=\sqrt{T_{//}/T_\perp}$ of 2 to 3 is desirable, which necessitates 4 or 6 silica plates. The 6-plate device is drawn in Fig. 1. Pure Herasil synthetic silica plates with dimensions 8.5 mm x 12 mm x 1.3 mm were cut and polished with a common wedge $\alpha$ of 12 arc minutes (3.5 mrad), sufficiently large for complete elimination of interferential birefringence, see below. The plates were cut with particular care to minimize stress birefringence[14]. The birefringence of each plate was measured at normal incidence between crossed polarizers with a laser at 633 nm, and was found to be $\Delta\phi \leq 2$ mrad, except near the edges. To minimize any additional stress birefringence, the plates are just laid on top of each other by their own weight, without being clamped or glued. During assembly the plates were tilted with respect to each other (see Fig. 1), so that no two surfaces were parallel. Moreover, a careful arrangement of the orientation of the plates ensures that the whole dichroic device does not translate or tilt the incident beam when inserted or removed from the laser path. The 6-plate device has a length of 23 mm, which is acceptable, and exhibits a net deviation of $\leq$ 1mrad.

For $N$ silica plates with $n = 1.45$ (visible and near IR range) at Brewster incidence $\theta_B = 55.4°$, one obtains $k = 1/(0.8737)^N$ (see Appendix B), while the transmission for the unattenuated polarization is $T_{//} = 1$. With the chosen value of $N = 6$, this leads to $k = 2.247$. Due to the requirement that no two surfaces be parallel, exact Brewster incidence is not possible for all plates and this results in losses. However, as seen in Fig. 2, even for the 6-plate magnifier, losses remain smaller than 3% in a range of several degrees. Actually, choosing an incidence angle slightly *above* Brewster incidence allows significant gain in the magnification factor $k$, at the expense of a reasonable loss in $T_{//}$. For this reason we chose an (average) angle of incidence of 60°, for which $k = 2.782$ and $T_{//} = 0.97$.

### III. MEASUREMENT OF THE MAGNIFYING FACTOR

Measurements of the magnification factor $k$ were made by using the two main methods of polarimetry, i.e., extinction between crossed polarizers and balanced mode, the latter being the polarimetric method used in our Parity Violation experiment[15]. In the first case, the device was inserted between crossed Glan polarizer and analyzer in the path of a laser beam at 633 nm. The *y*-axis of the device was set parallel to the polarizer transmitting axis. Then the device was tilted by $\psi \approx 1°$ around the laser beam and we measured the tilt $k\psi$ of the analyzer needed to restore extinction. We obtained $k = 2.85$ for the 6-plate device instead of 2.834 as calculated for 60° incidence and $n = 1.457$ at 633 nm. In the second polarimetric mode, the device was inserted in front of a balanced-mode two-channel polarimeter[16] operated with a 1.47 μm laser. From the ratio of the polarimeter imbalance, measured with and without the magnifier device and again for a tilt $\psi$ in the 1° range of the polarization incident on the 6-plate device, we obtained $k = 2.79$ while the expected value is 2.745 for $n = 1.445$ at 1.47µm. Thus good agreement is found between expected and measured values of $k$ and we have checked that this agreement also holds for a 4-plate device. Note that, in the case of our Parity Violation experiment[13,15], precise knowledge of the value of the magnification factor $k$ is not required since it is eliminated in the real-time calibration procedure.

## IV. CHARACTERIZATION OF THE BIREFRINGENCE

A helium-neon laser beam was used to check that the multiply reflected beams are tilted, with respect to the main beam, by the expected angle $2n\alpha \tan(\theta_B)$ equal to 15 mrad. This angle is much larger than the typical $\approx$ 1 mrad divergence of a laser beam and thus allows separation of the multiply reflected beams in the far field. In our polarimeter[16], we reject the spurious beams by focussing them with a lens of 100 mm focal length. In its focal plane, the beams appear as spots separated by 1500 µm and the main beam is isolated with a diaphragm of 400 µm diameter before impinging on the analyzer (Fig. 3). Our balanced mode polarimeter also allows us to measure the birefringence of the 6-plate device[10] and we have found: $\Delta\phi \leq 0.3°$. This is a real improvement as compared to the multilayer-coated device[10,17], where the birefringence changed by 4° or more for a 1° change of the angle of incidence. Here the incidence is not critical. Note that, with up to $\Delta\phi \approx 9°$ per plate at 60° incidence (see App. A), a six-plate device made of parallel plates would have yielded an unacceptable interferential birefringence.

## V. DISCUSSION

We do not address here the general issue of the improvement in the signal-to-noise ratio to be expected from the use of a polarization magnifier, because it depends critically on the noise characteristics of the signal to be measured[10]. In the case of our Parity Violation experiment the challenge is to measure to 1% precision a polarization rotation of the order of 1μrad on a pulsed infrared probe laser beam amplified by stimulated emission in a pulse-excited atomic cesium vapor. The polarization magnifier has been decisive in the substantial improvement we recently obtained[13], as compared to the 9% validation measurement[15]. Used in conjunction with our two-channel balanced mode polarimeter the device has allowed us to combine the advantages of near extinction polarimetry and balanced mode polarimetry. The noise on the Parity Violation measurement is close to the shot-noise limit, by no means a straightforward result in a pulsed pump-probe experiment.

With 6 wedged plates near Brewster incidence our device proves to be an efficient low birefringence polarization magnifier in which interferential birefringence has been completely suppressed. The birefringence, now limited by residual stress, would probably be further reduced by using larger plates. A higher value of the magnifying factor (see Fig. 4) could be obtained using a larger number of plates provided enough care were taken over the geometrical quality of the transmitted beam and the depolarization rate.

**ACKNOWLEDGMENTS**

We thank Loic Estève for his help in characterizing and checking of the wedged plates.

**APPENDIX A: Calculation of the interferential birefringence of a single parallel plate**

The amplitude of the transmitted field through a plate of index $n$ and thickness $e$ after an infinite number of internal reflections (the well-known Fabry-Perot formula) is

$$A = e^{i\varphi/2} t^{in} t^{out} / (1 - r^2 e^{i\varphi}) \qquad (A1)$$

where the phase $\varphi = \frac{2\pi}{\lambda} 2ne\cos(\theta_r)$ acquired after two internal reflections is the same for both polarizations parallel and perpendicular to the incidence plane. $\theta_r$ is the refracted angle. $t^{in}$, $t^{out}$ are the transmission amplitude coefficients for the air-to-silica and silica-to-air interfaces respectively, and r the reflection amplitude coefficient at the silica-to-air interface. Introducing $t_{//} = t_{//}^{in} t_{//}^{out}$, $t_\perp = t_\perp^{in} t_\perp^{out}$ and $R_{//} = r_{//}^2$, $R_\perp = r_\perp^2$ the transmitted amplitudes for the parallel and perpendicular polarizations are

$$A_{//} = e^{i\varphi/2} t_{//} / (1 - R_{//} e^{i\varphi}) = c_{//} e^{i\phi_{//}} \qquad (A2a)$$

$$A_\perp = e^{i\varphi/2} t_\perp / (1 - R_\perp e^{i\varphi}) = c_\perp e^{i\phi_\perp}, \qquad (A2b)$$

respectively, where $c_{//}$ and $c_\perp$ are real coefficients. Then the phase difference $\Delta\phi = \phi_{//} - \phi_\perp$ between the two transmitted polarizations is simply extracted from the ratio $A_{//}/A_\perp$. We find

$$\tan(\Delta\phi) = \tan(\phi_{//} - \phi_\perp) = \frac{(R_{//} - R_\perp)\sin(\varphi)}{1 + R_{//} R_\perp - (R_{//} + R_\perp)\cos(\varphi)}, \qquad (A3)$$

where the reflection coefficients are given by

$$R_{//} = \left(\frac{\tan(\theta - \theta_r)}{\tan(\theta + \theta_r)}\right)^2 \text{ and} \qquad (A4a)$$

$$R_\perp = \left(-\frac{\sin(\theta - \theta_r)}{\sin(\theta + \theta_r)}\right)^2, \qquad (A4b)$$

using, for instance, eqs.(21a) in Ref.[18], §1.5.2, for the Fresnel formulae at an incidence angle $\theta$. At Brewster incidence, $\theta + \theta_r = 90°$, $R_{//} = 0$, hence $\tan(\Delta\phi) = \frac{-R_\perp \sin(\varphi)}{1 - R_\perp \cos(\varphi)}$ whose maximum value, obtained for $\varphi = -90°$, is $\tan(\Delta\phi_{max}) = R_\perp$. For silica with n =1.45, $\theta_B = 55.4°$, hence $R_\perp = 0.126$ and $\Delta\phi_{max} = 7.2°$. For the incidence angle of 60° chosen here, we find $R_\perp = 0.158$ and $\Delta\phi_{max} = 9°$.

## APPENDIX B: Calculation of the magnification factor for a single plate

We use formulae (20a) from Ref.[18], §1.5.2, to express the transmission *amplitude* coefficients as

- for the air to silica interface:

$$t_{//}^{in} = 2 \cos\theta \sin\theta_r / \left[\sin(\theta+\theta_r) \cos(\theta-\theta_r)\right] \quad \text{(B1a)}$$

$$t_{\perp}^{in} = 2 \cos\theta \sin\theta_r / \sin(\theta+\theta_r) \quad \text{(B1b)}$$

where $\theta$ and $\theta_r$ are the incident and refracted angles,

- for the silica to air interface:

$$t_{//}^{out} = 2 \cos\theta_r \sin\theta / \left[\sin(\theta+\theta_r) \cos(\theta-\theta_r)\right] \quad \text{(B2a)}$$

$$t_{\perp}^{out} = 2 \cos\theta_r \sin\theta / \sin(\theta+\theta_r). \quad \text{(B2b)}$$

Then, for a single plate, the relation $k = t_{//}^{in} t_{//}^{out} / t_{\perp}^{in} t_{\perp}^{out} = 1 / \cos^2(\theta-\theta_r)$ immediately follows, for any angle of incidence $\theta$.

# REFERENCES

[a] E-mail: lintz@lkb.ens.fr

[b] Present address: Laboratoire de Physique des Lasers, Université Rennes 1, Campus de Beaulieu, F-35042 Rennes cedex, France

[c] Laboratoire de l'Université Pierre et Marie Curie et de l'Ecole Normale Supérieure, associé au CNRS (UMR 8552)

[d] Unité FR 684 du CNRS

[1] M.-A. Bouchiat and C. Bouchiat, Rep. Prog. Phys. **60**, 11 (1997).

[2] M. J. D. Macpherson, K. P. Zetie, R.B. Warrington, D. N. Stacey, J. P. Hoare, Phys. Rev. Lett. **67,** 2784 (1991).

[3] D. M. Meekhof, P. Vetter, P. K. Majumder, S. K. Lamoreaux, E. N. Fortson, Phys. Rev. Lett. **71**, 3442 (1993).

[4] N. H. Edwards, S. J. Phipp, P. E. G. Baird, S. Nakayama, Phys. Rev. Lett. **74**, 2654 (1995).

[5] D. Bakalov, F. Brandi, G. Cantatore, G. Carugno, S. Carusotto, F. Della-Valle, A.-M. De-Riva, U. Gastaldi, E. Iacopini, P. Micossi, E. Milotti, R. Onofrio, R. Pengo, F. Perrone, G. Petrucci, E. Polacco, C. Rizzo, G. Ruoso, E. Zavattini, G. Zavattini, Quantum Semiclass. Opt. **10**, 239 (1998).

[6] I.K. Kominis, T. W. Kornack, J. C. Allred, M. V. Romalis, Nature **422**, 596 (2003).

[7] D. Budker, W. Gawlik, D. F. Kimball, S. M. Rochester, V. V. Yashchuk, A. Weis, Rev. Mod. Phys. **74**, 1153 (2002),

[8] D. Budker, D. F. Kimball, S. M. Rochester, V. V. Yashchuk, M. Zolotorev, Phys. Rev. **A 62**, 043403 (2000).

[9] R. M. A. Azzam, N. M. Bashara, "Ellipsometry and polarized light" (North-Holland, Amsterdam, 1977).

[10] D. Chauvat, J. Guéna, Ph. Jacquier, M. Lintz, M.-A. Bouchiat, M. D. Plimmer, C. W. Goodwin, Opt. Commun. **138**, 249 (1997).

[11] V. S. Zapasskii, Opt. Spectr. **47**, 450 (1979).

[12] E. Collett, Appl. Opt. **11**, 1184 (1972).

[13] J. Guéna, M. Lintz, M.-A. Bouchiat, arXiv : physics/0412017.

[14] Crystal Laser International, 6 rue Pont Vieux, F30460 Lasalle, France. Tel (33) 4 66 34 09 39.

[15] J. Guéna, D. Chauvat, Ph. Jacquier, E. Jahier, M. Lintz, S. Sanguinetti, A. Wasan, M.-A. Bouchiat, A. V. Papoyan, D. Sarkisyan, Phys. Rev. Lett. **90**, 143001 (2003).

[16] J. Guéna, Ph. Jacquier, M. Lintz, L. Pottier, M.-A. Bouchiat, A. Hrisoho, Opt. Commun. **71**, 6 (1989).

[17] S. Sanguinetti, Ph. D. Thesis (Université Pierre et Marie Curie and Universita di Pisa, 2004, http://tel.ccsd.cnrs.fr/documents/archives0/00/00/67/85/ ), sect. 2-4-3.

[18] M. Born and E. Wolf: "Principle of Optics", second edition (Pergamon Press, Oxford, 1964).

**Fig 1**

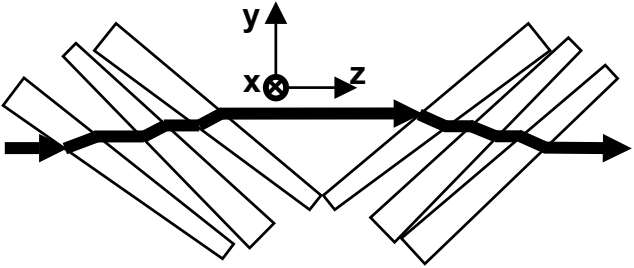

*FIG. 1: Arrangement chosen for the 6-plate magnifier, in which no two surfaces are parallel (the wedge and tilts of the plates are exaggerated).*

**Fig 2**

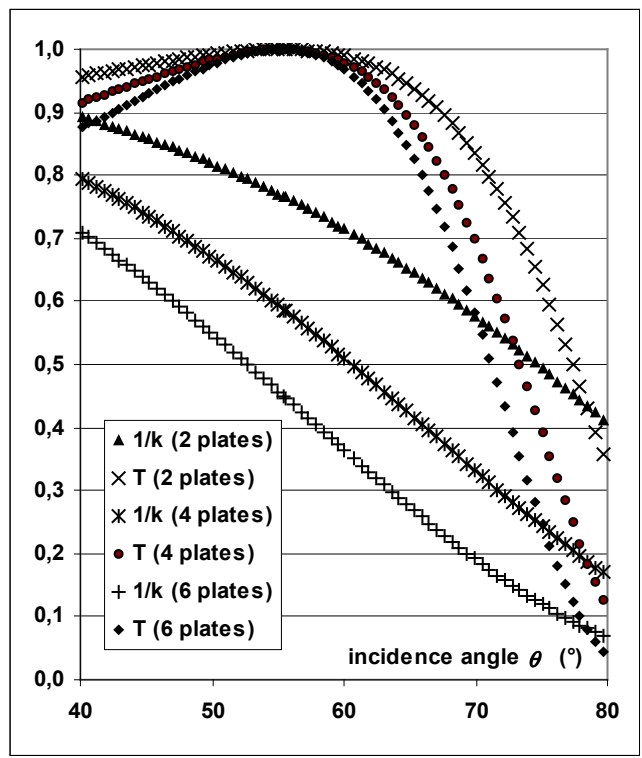

*FIG. 2: Calculated variations of the total intensity transmission $T_{//}$ and $1/k = \sqrt{T_\perp / T_{//}}$ (instead of k, for convenience) as a function of the incidence angle. Index of the plate material: n=1.45. For simplification, the incidence angle is the same for all plates and the plates are assumed to be parallel (but interference effects are supposed not to take place).*

**Fig 3**

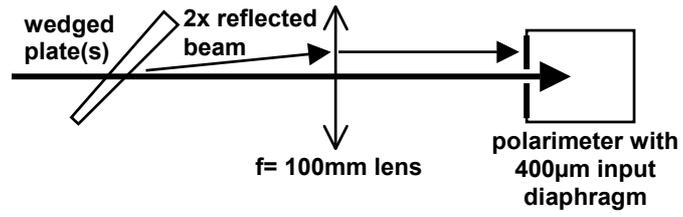

*FIG. 3: Rejection of the multiply-reflected beams generated by the polarization magnifier.*

**Fig 4**

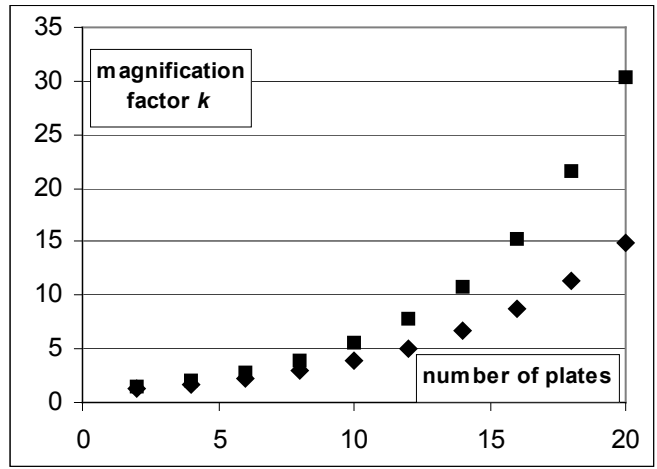

*FIG. 4: Calculated magnification factor k vs number N of silica plates. Diamonds: for plates at Brewster incidence. Squares: plates at 60° incidence.*